 \documentclass[manuscript]{aastex61}

\usepackage{color}
\usepackage{ulem}
\usepackage{xcolor}
\usepackage{soul}
\usepackage{CJKutf8}
\newcommand{\cntext}[1]{\begin{CJK}{UTF8}{gbsn}#1\end{CJK}\kern-1ex}

\begin{document}

\title{The Acceleration and Confinement of Energetic Electrons by a Termination Shock in a Magnetic Trap: An Explanation for Nonthermal Loop-top Sources during Solar Flares}


\author[0000-0003-1034-5857]{Xiangliang Kong  (\cntext{孔祥良})}
\affil{Shandong Provincial Key Laboratory of Optical Astronomy and Solar-Terrestrial Environment,
and Institute of Space Sciences, Shandong University, Weihai, Shandong 264209, People's Republic of China; kongx@sdu.edu.cn}
\affiliation{Sate Key Laboratory of Space Weather, Chinese Academy of Sciences, Beijing 100190, People's Republic of China}

\author[0000-0003-4315-3755]{Fan Guo  (\cntext{郭帆})}
\affiliation{Los Alamos National Laboratory, Los Alamos, NM 87545, USA}
\affiliation{New Mexico Consortium, 4200 West Jemez Rd, Los Alamos, NM 87544, USA}

\author[0000-0002-9258-4490]{Chengcai Shen  (\cntext{沈呈彩})}
\affiliation{Harvard-Smithsonian Center for Astrophysics, 60 Garden St, Cambridge, MA 02138, USA}

\author[0000-0002-0660-3350]{Bin Chen (\cntext{陈彬})}
\affiliation{Center for Solar-Terrestrial Research, New Jersey Institute of Technology,
323 Dr. Martin Luther King Blvd, Newark, NJ 07102, USA}

\author[0000-0001-6449-8838]{Yao Chen  (\cntext{陈耀})}
\affil{Shandong Provincial Key Laboratory of Optical Astronomy and Solar-Terrestrial Environment,
and Institute of Space Sciences, Shandong University, Weihai, Shandong 264209, People's Republic of China; kongx@sdu.edu.cn}

\author{Sophie Musset}
\affil{School of Physics and Astronomy, University of Minnesota, Minneapolis, MN 55455, USA}

\author{Lindsay Glesener}
\affil{School of Physics and Astronomy, University of Minnesota, Minneapolis, MN 55455, USA}

\author{Peera Pongkitiwanichakul}
\affil{Department of Physics, Faculty of Science, Kasetsart University, Bangkok 10900, Thailand}

\author{Joe Giacalone}
\affiliation{Department of Planetary Sciences, University of Arizona, Tucson, AZ 85721, USA}

\begin{abstract}
Nonthermal loop-top sources in solar flares are the most prominent observational signature that suggests energy release and particle acceleration in the solar corona. Although several scenarios for particle acceleration have been proposed, the origin of the loop-top sources remains unclear. Here we present a model that combines a large-scale magnetohydrodynamic simulation of a two-ribbon flare with a particle acceleration and transport model for investigating electron acceleration by a fast-mode termination shock at the looptop. Our model provides spatially resolved electron distribution that evolves in response to the dynamic flare geometry. We find a concave-downward magnetic structure located below the flare termination shock, induced by the fast reconnection downflows. It acts as a magnetic trap to confine the electrons at the looptop for an extended period of time. The electrons are energized significantly as they cross the shock front, and eventually build up a power-law energy spectrum extending to hundreds of keV. We suggest that this particle acceleration and transport scenario driven by a flare termination shock is a viable interpretation for the observed nonthermal loop-top sources.

\end{abstract}


\section{Introduction} \label{sec:intro}
Solar flares are the most powerful energy release phenomena and important sites for particle acceleration in the solar system \citep{benz17}. It is believed that the stored magnetic energy explosively releases via magnetic reconnection \citep{shibata11}.
Observations have shown that a significant fraction of the released magnetic energy goes into the accelerated nonthermal particles \citep[e.g.,][]{emslie12,aschwanden17}.
However, it remains controversial how a large number of particles ($>$10$^{36}$ electrons) are impulsively accelerated to high energies within tens of seconds to minutes \citep{miller97}. Proposed acceleration mechanisms include acceleration in the reconnection layers \citep{drake06,oka10,li18a,li18b}, stochastic acceleration by plasma turbulence \citep{miller96,lazarian99,petrosian04,pongkitiwanichakul14}, and shock acceleration in the reconnection outflows \citep{mann09,guo12,li13a}.

The most remarkable evidence for flare particle acceleration is hard X-ray (HXR) emission source above the top of soft X-ray loops first reported by \citet{masuda94}. Similar events with high-energy emissions have been observed since then \citep[e.g.,][]{Melnikov02,krucker10,liu13,krucker14,oka15,gary18}.
The loop-top or above-the-loop-top (referred to as ``loop-top" hereafter) HXR and microwave sources indicate energy release and particle acceleration in the corona with energetic electrons typically taking a power-law energy distribution.
In addition, the confinement of energetic electrons at the loop top is another  issue for the formation of coronal HXR sources. \citet{simos13} found that the number of energetic electrons required to explain observations is 2--8 times higher at the looptop than at the footpoints, indicating electron trapping at the loop top. A successful model for the particle acceleration and transport in solar flares must be able to explain these important observations for the loop-top sources.

A fast-mode termination shock (TS) has been proposed to explain electron energization at the loop-top region \citep[][]{masuda94,tsuneta98}. The formation of a TS has also been predicted in numerical magnetohydrodynamic (MHD) simulations when high-speed reconnection outflows impinge on magnetic loops \citep[e.g.,][]{forbes86,magara96,yokoyama98,yokoyama01,takasao15,shen18}.
However, the existence of a TS and its role in particle acceleration have remained uncertain due to the lack of evidence in observations. Recently, using the Karl G. Jansky Very Large Array, \citet{chen15} revealed the presence of a TS in an eruptive flare based on radio spectroscopic imaging and provided observational evidence for its role in accelerating electrons to at least tens of keV. They showed that the TS, which manifests as stochastic radio spikes, was located at the front of reconnection downflows and was slightly above the loop-top HXR source.

Despite promising development in observations and theories of loop-top sources, to our best knowledge no study has successfully modeled the spatial distribution of accelerated electrons for explaining high-energy emissions at the loop-top region.
As discussed above, this may require both acceleration and confinement of electrons at the loop-top.
Some models such as magnetic mirroring, turbulent pitch-angle scattering, and formaiton of thermal fronts have been suggested to explain the confinement of energetic electrons within the flare loops \citep[e.g.,][]{li13b,simos13,kontar14,musset18,sun19}.
However, it remains unclear how energetic electrons are confined at the looptop region.
In previous models, the TS is often considered as a planar standing shock \citep[e.g.,][]{tsuneta98,mann09}.
Recent MHD simulations have revealed that the TS can be very dynamic and have complex structures \citep[e.g.,][]{takasao15,takasao16,takahashi17,shen18}.
The magnetic structures in the TS region may affect the acceleration and transport of particles near the loop-top region. Especially, a concave-downward magnetic structure is found below the termination shock, which may trap electrons at the looptop, as it is more difficult for particles to travel transverse to the magnetic field than along it \citep{guo10,kong15,kong16}.
Although such a magnetic configuration has been shown in earlier MHD simulations \citep[e.g.,][]{magara96}, its role in confining energetic electrons has not been investigated heretofore.

In this study, we numerically model the acceleration of energetic electrons at the flare TS based on MHD simulations of a two-ribbon solar flare and emphasize the importance of a concave-downward magnetic trap structure in the TS region as a confinement mechanism for the formation of loop-top HXR sources. We find that the accelerated electrons are concentrated in the loop-top region due to acceleration at the TS and confinement by the magnetic trap structure. As far as we know, this is the first model that reproduces the necessary electron acceleration and spatial distribution for the loop-top sources. Section 2 introduces our MHD simulations and particle acceleration and transport model. In Section 3, we present the simulation results, and examine the electron acceleration and confinement. In Section 4, we discuss the conclusions and implications of this work.

\section{Numerical Methods} \label{sec:model}
\subsection{MHD simulations of the TS}

We model the reconnection-driven TS in a classic two-ribbon flare configuration by solving the two-dimensional resistive MHD equations using the Athena MHD code \citep{stone08}.
The initial setup is a vertical Harris-type current sheet along the $y$-direction in mechanical and thermal equilibrium. We employ an initial magnetic field perturbation to speed up the reconnection onset.
We use line-tied boundary condition at the bottom and open conditions at other boundaries so the magnetic field and plasma can move in or out the simulation domain freely.
The heat conduction is not included and the specific heat ratio is 5/3.
We use a uniform resistivity that corresponds to a Lundquist number $S$ = 10$^5$.
The simulation box is $x$ = [-1, 1] and $y$ = [0, 2], and 2048 $\times$ 2048 Cartesian grids are uniformly spaced.
The simulations are normalized by the length $L_0$ = 75 Mm, the velocity $V_0$ = 810 km s$^{-1}$, and the time $t_0$ = 92 s.
The parameters and setup in the MHD simulation are the same as Case 1 in \citet{shen18}.

\subsection{Modeling electron acceleration by solving the Parker transport equation}

In this study we focus on a period ($96.5 - 97.5 t_0$) where the termination shock is nearly steady and symmetric \citep{shen18}. The acceleration and transport process during dynamical evolution phase of the shock will be discussed in a future publication.  We select the region $x$ = [-0.25, 0.25] and $y$ = [0.2, 0.7] as the simulation domain for particle acceleration.

Electron acceleration and transport are modeled by numerically solving the Parker's transport equation \citep{parker65},
\begin{equation}  
\frac{\partial f}{\partial t} =
\frac{\partial}{\partial x_i} \left[ \kappa_{ij} \frac{\partial f}{\partial x_j} \right]
- U_i \frac{\partial f}{\partial x_i}
+ \frac{p}{3} \frac{\partial U_i}{\partial x_i} \frac{\partial f}{\partial p}
+ Q,
\end{equation} 
where $f(x_i, p, t)$ is the particle distribution function dependent on position $x_i$,  momentum $p$, and time $t$;
$\kappa_{ij}$ is the spatial diffusion coefficient tensor, $U_i$ is the bulk plasma velocity, and $Q$ is the source. The equation is solved by particle-based stochastic differential equations \citep{giacalone08,guo10,kong17,kong19,li18b}, and the fluid velocity and magnetic field necessary for solving the equations are from MHD simulations.
The temporal cadence of MHD frames is 0.01 $t_0$ and no interpolation is applied in time, meaning that we assume a steady TS between adjacent MHD frames.
We use a bilinear interpolation in space to deduce the physical quantities at the particle position.

The spatial diffusion coefficient describes particle transport in the magnetic field.
The diffusion coefficient tensor is,
\begin{equation}
\kappa_{ij} = \kappa_{\perp} \delta_{ij} + (\kappa_{\parallel} - \kappa_{\perp}) \frac{B_i B_j}{B^2},
\end{equation}
where $\kappa_{\parallel}$ and $\kappa_{\perp}$ are the parallel and perpendicular diffusion coefficients, and $B_{i}$ is the average magnetic field vector.
The antisymmetric diffusion coefficient related to particle drifts is neglected because the gradient and curvature drifts are in the out-of-plane direction. 
$\kappa_{\parallel}$ can be calculated from the quasilinear theory  \citep{jokipii71}.
We assume the magnetic turbulence is well developed and has a Kolmogorov power spectrum $P \propto k^{-5/3}$, then the resulting diffusion coefficient $\kappa_{\parallel} \propto p^{4/3}$
when the particle gyroradius is much smaller than the turbulence correlation length.
We use the following expression for $\kappa_{\parallel}$ \citep{giacalone99},
\begin{equation}
\kappa_{\parallel} = \frac{3 v^3}{20 L_c \Omega^2 \sigma^2}
csc\left(\frac{3 \pi}{5}\right) \left[1 + \frac{72}{7} \left(\frac{\Omega L_c}{v} \right)^{5/3} \right],
\end{equation}  
where $v$ is the particle speed, $L_c$ is the turbulence correlation length, $\sigma^2$ is the normalized wave variance of turbulence, and $\Omega$ is the particle gyrofrequency.
The normalization of the diffusion coefficient is $\kappa _0$ = $L_0 V_0$ = 6 $\times 10^{17}$ cm$^2$ s$^{-1}$.
We assume the turbulence correlation length $L_c$ = 1 Mm, the average magnetic field $B$ = 100 G, and the normalized wave variance of turbulence $\sigma^2 = <\delta B^2> / B_0^2$ = 0.6. Then, $\kappa_{\parallel 0} = 3 \times 10^{15}$ cm$^2$ s$^{-1}$ for the electron initial energy $E_0$ = 5 keV, corresponding to 0.005 $\kappa_0$.
Here we take $\kappa_{\perp} / \kappa_\parallel$ = 0.1 similar to results of test-particle simulations in synthetic turbulence \citep{giacalone99}. Other kinetic processes that may increase the pitch-angle scattering rate is not included in the simple model.
As shown in \citet{shen18}, the TS can be resolved in several cells in MHD simulations, which means that the shock width is on the order of one cell, $\sim$0.001 $L_0$. Therefore the characteristic diffusion length at the shock $L_{d} = \kappa_{nn} /V_{sh}$, where $\kappa_{nn}$ is the diffusion coefficient in the shock normal direction, is roughly the shock width for a quasi-perpendicular TS. With these parameters, the electron energy distribution resembles a power-law shape, close to that predicted by the diffusive shock acceleration (DSA) theory.

\section{Simulation Results} \label{sec:result}

Figures 1(a) and (b) show the distributions of $y$-component of plasma flow velocity $V_y$ and the divergence of plasma velocity $\nabla \cdot \textbf{V}$ at time 96.5 $t_0$, respectively.
A TS forms when the fast reconnection outflow collides with closed loops and can be well manifested by negative $\nabla \cdot \textbf{V}$ regions located at the front of reconnection outflow, as marked by the arrow in Figure 1(b).
The TS is very dynamic, with the morphology and physical quantities varying in space and time \citep{shen18}. Figures 1(c)-(e) show the temporal variations of the maximum and average values of fast Mach number $M_F$, density compression ratio $X$, and shock angle $\theta_{Bn}$ at the TS front from 95 $t_0$ to 100 $t_0$, respectively.
Consistent with the observational results based on the split-band feature of the TS-associated stochastic spike bursts \citep{chen19}, the maximum (average) $M_F$ ranges from 1.5 to 3.2 (1.4 to 2.4), the maximum (average) $X$ ranges from 1.6 to 3.7 (1.4 to 2.5), and the maximum (average) $\theta_{Bn}$ ranges from 66$^{\circ}$ to 90$^{\circ}$ (12$^{\circ}$ to 82$^{\circ}$).
Here we focus on a relatively steady period between 96.5 $t_0$ and 97.5 $t_0$ and utilize the plasma velocity and magnetic field to perform particle acceleration modeling.
During this period, the average density compression ratio $X$ is $\sim$2 and shock angle $\theta_{Bn}$ is 60$^{\circ}$-70$^{\circ}$.

In the particle acceleration simulation, a total of 2.4 $\times 10^{6}$ pseudo-particles are injected at a constant rate upstream of the TS. The particles have the same initial energy of 5 keV. To improve statistics at high energies, we have implemented a particle-splitting technique so a pseudo-particle will be splitted into more particles at higher energy \citep{kong17,kong19}.
Particle acceleration at quasi-perpendicular shocks are usually considered as the so-called shock drift acceleration \citep{wu84,mann09}. In a diffusion approximation, the shock drift acceleration can be included in the DSA \citep{jokipii87}.
For $X \approx 2$, the DSA predicts a power-law distribution in momentum $f(p) \propto p^{-3X/(X-1)} = p^{-6}$. Therefore the differential distribution in energy has the form $dJ/dE = p^2f \propto E^{- \delta}$ with the spectral index $\delta = (X+2)/[2(X-1)] \approx 2$ in the non-relativistic limit.

Figure 2(a) shows the temporal variations of energy spectra of accelerated electrons integrated over the whole simulation domain.
The low-energy spectra below 50 keV are approximately power-law distributions with a spectral index $\delta$ = 2.5, close to the prediction from 1-D steady state DSA solution. With the thin-target emission model, the photon power-law index is $\gamma_{thin} = \delta + 1 \approx 3.5$, which is generally harder than that deduced from loop-top HXR sources \citep{oka18}.
As in the DSA the power-law index depends strongly on the compression ratio, a TS with a smaller compression ratio (e.g., $X \sim$ 1.5) may produce electron spectra consistent with observations ($\delta \approx 3.5$, then $\gamma_{thin} \approx 4.5$). The compression ratio can be affected by various parameters such as the plasma $\beta$, guide field, and thermal conduction in MHD simulations \citep[e.g.,][]{seaton09,takasao16,shen18}. As noted in \citet{oka18}, in some flares the loop-top HXR emission can extend to the $\gamma$-ray range but shows hard spectra close to our simulation results in the X-ray range below 100 keV \citep[e.g.,][]{Pesce-Rollins15}.
Figure 2(b) shows the number of electrons accelerated to different energies as a function of time. Electrons can be accelerated to $100$ keV in a few seconds.
In DSA the acceleration time depends mainly on the particle diffusion coefficient. The smaller the diffusion coefficient, the higher energy can be obtained for a given time.

Figure 3(a) shows the distribution of $\nabla \cdot \textbf{V}$ at 97.5 $t_0$, when the TS is at $y$ $\sim$0.6 $L_0$. Magnetic field lines in red illustrate the magnetic trap structure.
Figures 3(b)-(d) show the spatial distributions of accelerated electrons at 97.5 $t_0$ for three different energy ranges, 5--10 keV, 20--30 keV, and 50--100 keV, respectively.
In all energy ranges, a loop-top source, where energetic electrons are concentrated, appears below the TS and above closed loops, coinciding with the configuration of the magnetic trap structure and consistent with the observations in \citet{chen15}.
It also shows that the higher the electron energy, the source size is smaller and the source is located closer to the TS (in higher altitude). The loop-top sources for 20--30 keV and 50--100 keV are located $\sim$0.1 $L_0$ above closed loops, corresponding to $\sim$7 Mm or 10$''$.
Those results are consistent with observations of loop-top HXR sources \citep[][]{liu08,liu13,krucker10,oka15}.
Note that in observations the dependence of source size with energy for coronal HXR sources within the loop remains controversial \citep{dennis18}.
We also found that the loop-top source does not qualitatively change when we vary $\kappa_{\perp} / \kappa_\parallel$. However, it does quantitatively control the efficiency of the confinement and the flux contrast between the loop-top source and rest of the simulation region. More detailed model-observation comparisons are required to test this model and its parameter dependence.

To further illustrate the effect of the magnetic trap structure, Figure 4 shows the trajectory of a representative pseudo-particle.
The particle moves roughly along the large-scale field lines and travels back and forth as seen in the $x$ position.
It is accelerated mainly at the TS and has three rapid acceleration phases as highlighted in orange, magenta, and green, respectively.
After the second acceleration phase, the electron energy has reached 70 keV and is trapped in the concave-downward magnetic structure. Later it moves upward and is guided back to the TS and receives a further acceleration to 120 keV. Finally the electron escapes from the magnetic trap. The electron can encounter the TS for multiple times via large-scale field lines, because the downstream flow can be very slow in the shock frame and the electron can cross field lines due to perpendicular diffusion.
This suggests that the concave-downward magnetic trap structure not only can confine electrons but also help particle acceleration.

Overall, the simulation results show a concentrated nonthermal electron source at the loop top similar to what is expected from HXR and microwave observations. Figure 5 shows our picture for explaining the loop-top source by including a concave-downward magnetic trap structure in the loop-top region based on our numerical simulations. As in the standard solar flare model, magnetic reconnection outflow collides with plasma at the loop-top region and creates a TS. Because of this process, the newly reconnected field lines and field lines at the loop-top region form a magnetic trap structure that confines energetic electrons. Interestingly, the magnetic trap is not completely closed, and electrons are still allowed to escape from the loop-top region along some field lines connecting to the footpoints.

\section{Conclusions and Discussion} \label{sec:conclusion}

In this paper, we present a MHD--particle model for studying particle acceleration and transport by a TS in a two-ribbon solar flare. For a set of parameters reasonable for solar flares, we obtain fast electron acceleration into a power-law distribution to up to several hundred keV. The accelerated electrons are confined in the loop-top region due to a concave-downward magnetic structure, which we refer to as a magnetic trap, forming a concentrated source as expected for producing high-energy emission in the corona.  For the first time, we are able to reproduce energetic electron energy and spatial distributions that are necessary for explaining the loop-top nonthermal sources.

This model can readily be coupled with radiation models for calculating  HXR and microwave emissions and add more realistic effects such as Coulomb collision. We note that the particle mean free path in our simulation $\lambda_{\parallel} = 3 \kappa_{\parallel}/v \approx$ 20 km for 5 keV electrons, which is smaller than that deduced in some flares \citep[e.g.,][]{kontar14,musset18}.
However, the turbulence properties and electron mean free path at the shock region remain unknown. To achieve efficient acceleration to hundreds of keV, our model requires the diffusion coefficient to be relatively small at the shock. Using a larger diffusion coefficient will reduce the shock acceleration rate. It is worthwhile noting that other nondiffusive acceleration processes may happen but not included in the DSA \citep{tsuneta98,jokipii07,guo10a,guo12}, so the acceleration results obtained here may only be a lower limit, or can be achieved by a lower turbulence amplitude assumed in the simulation.

The simulations presented in this paper are for a quasi-steady and nearly symmetric phase of the TS evolution. If the TS has an important contribution to particle acceleration in solar flares, we expect that the dynamical evolution of the TS is important in modulating the acceleration of particles and the associated high-energy emissions. One piece of such observational evidence was provided by \citet{chen15}, who showed that a temporary disruption of the termination shock front, revealed by radio dynamic spectroscopic imaging, coincided with a reduction of the nonthermal radio and HXR emission. Another outstanding example is the quasi-periodic pulsations (QPPs), which may be associated with modulations of the particle acceleration processes, among others (see, e.g., reviews by \citealt{nakariakov09} and \citealt{McLaughlin18}). One of the physical scenarios, as proposed by \citet{takasao16}, attributes the QPPs to the spontaneous quasi-periodic oscillation of multiple dynamic shocks (including the fast-mode termination shock) in the looptop region. In the future, we will study how the evolution of the TS influences particle acceleration and distribution in the loop-top region spatially and temporally.
We also note that our current MHD model does not include the effect of heat conduction. The introduction of heat conduction, as shown in previous studies \citep{seaton09,takasao16}, can affect the formation, geometry, and dynamics of the termination shock(s) and, in turn, the associated electron acceleration. It is hence of particular interest to investigate the effects of heat conduction on electron acceleration and transport by the flare termination shocks in the future.

To conclude, this work may have a strong implication to high-energy solar flare studies, as future development of this technique may enable detailed comparison between particle acceleration model and HXR and microwave observations in both space and time.

\acknowledgments

This work was supported by the National Natural Science Foundation of China under grants 11873036 and 11790303 (11790300), the Young Elite Scientists Sponsorship Program by China Association for Science and Technology, the Young Scholars Program of Shandong University, Weihai, and the Project Supported by the Specialized Research Fund for State Key Laboratories.
F.G. acknowledges the support from the National Science Foundation under grant 1735414 and support from by the U.S. Department of Energy, Office of Science, Office of Fusion Energy Science, under Award Number DE-SC0018240.
B.C. acknowledges the support from the National Science Foundation under grants AGS-1654382 and AST-1735405.
P.P. acknowledges support from Thailand Science Research and Innovation through the grant RTA6280002.
J.G. acknowledges support from the National Science Foundation under grants 1735422 and 1931252.
The work was carried out at National Supercomputer Center in Tianjin (TianHe-1A) and National Supercomputer Center in Guangzhou (TianHe-2).



\begin{figure}
\centering
\includegraphics[width=0.55\linewidth]{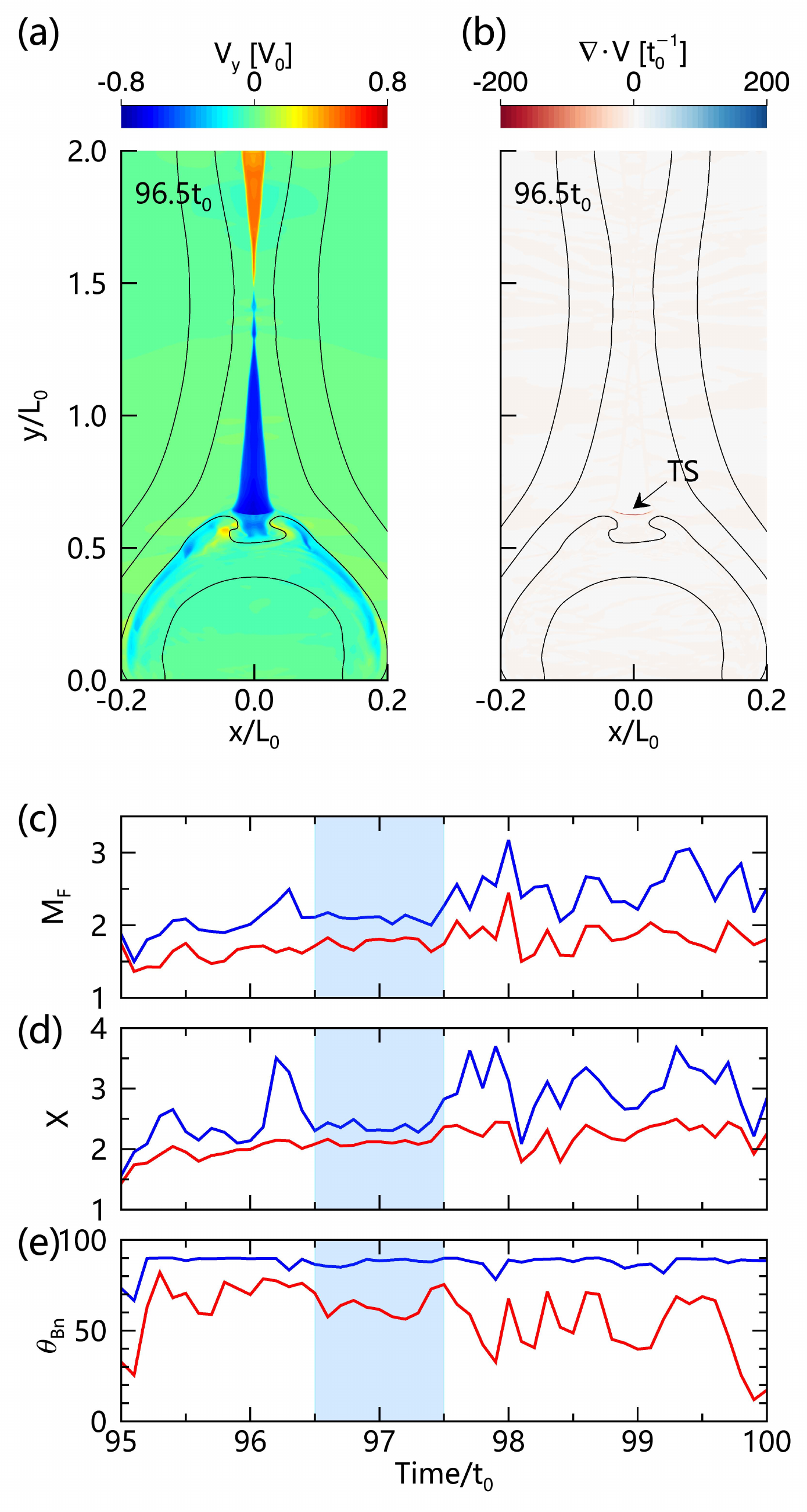}
\caption{
MHD modeling of the TS. (a) and (b) Distributions of $y$-component of plasma velocity $V_y$ and the divergence of plasma velocity $\nabla \cdot \textbf{V}$ at 96.5 $t_0$. The black lines denote the magnetic field.
(c)--(e) Temporal variations of the maximum (blue lines) and
average (red lines) of fast Mach number $M_F$, density compression ratio $X$, and shock angle $\theta_{Bn}$ at the TS front from 95 $t_0$ to 100 $t_0$. The shaded period is selected to perform particle simulation.
}
\label{fig:mhd}
\end{figure}

\begin{figure}
\centering
\includegraphics[width=0.6\linewidth]{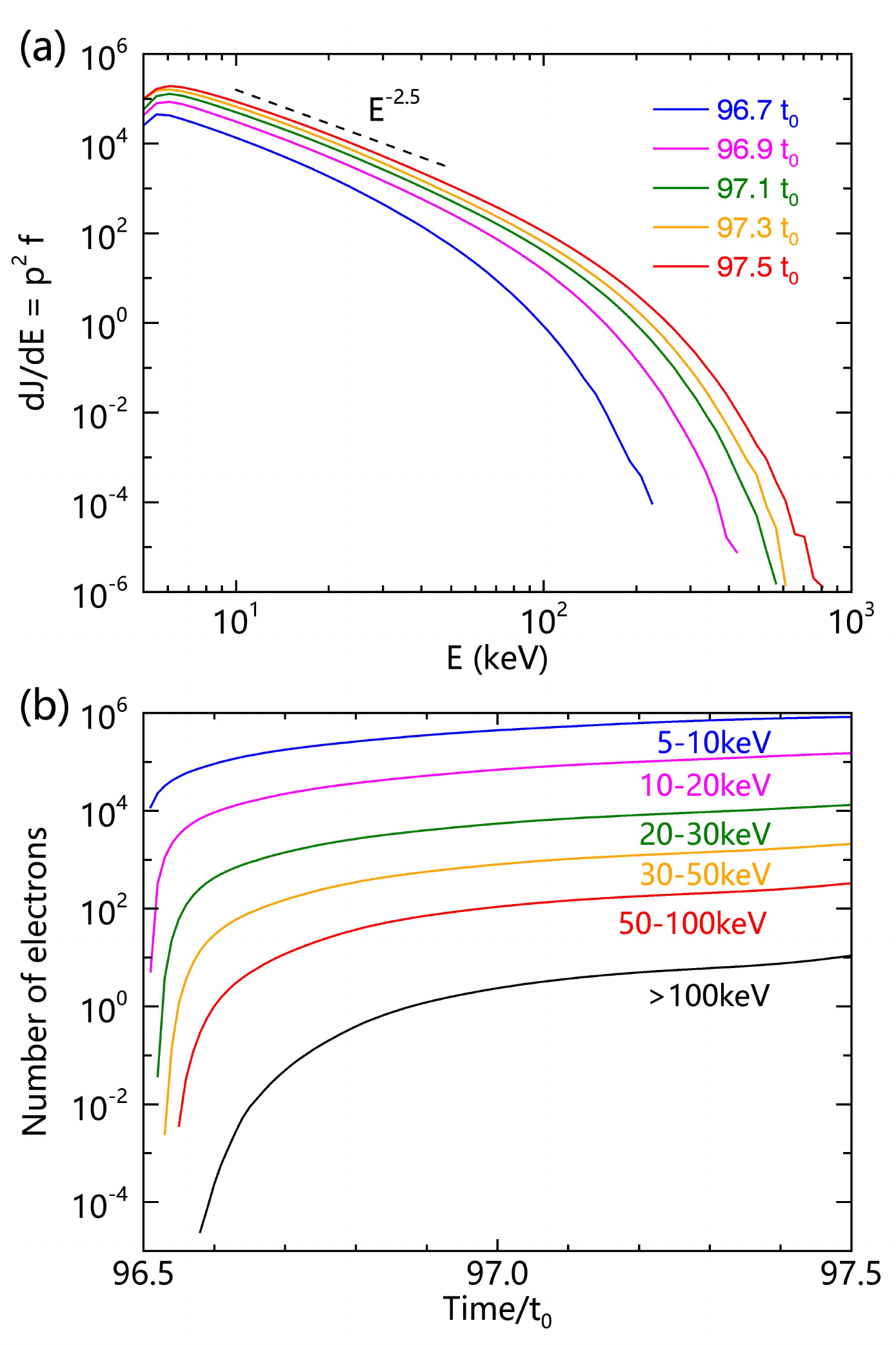}
\caption{
Temporal variations of (a) energy spectra of accelerated electrons and (b) number of electrons at different energy ranges.
}
\label{fig:spectrum}
\end{figure}

\begin{figure}
\centering
\includegraphics[width=0.85\linewidth]{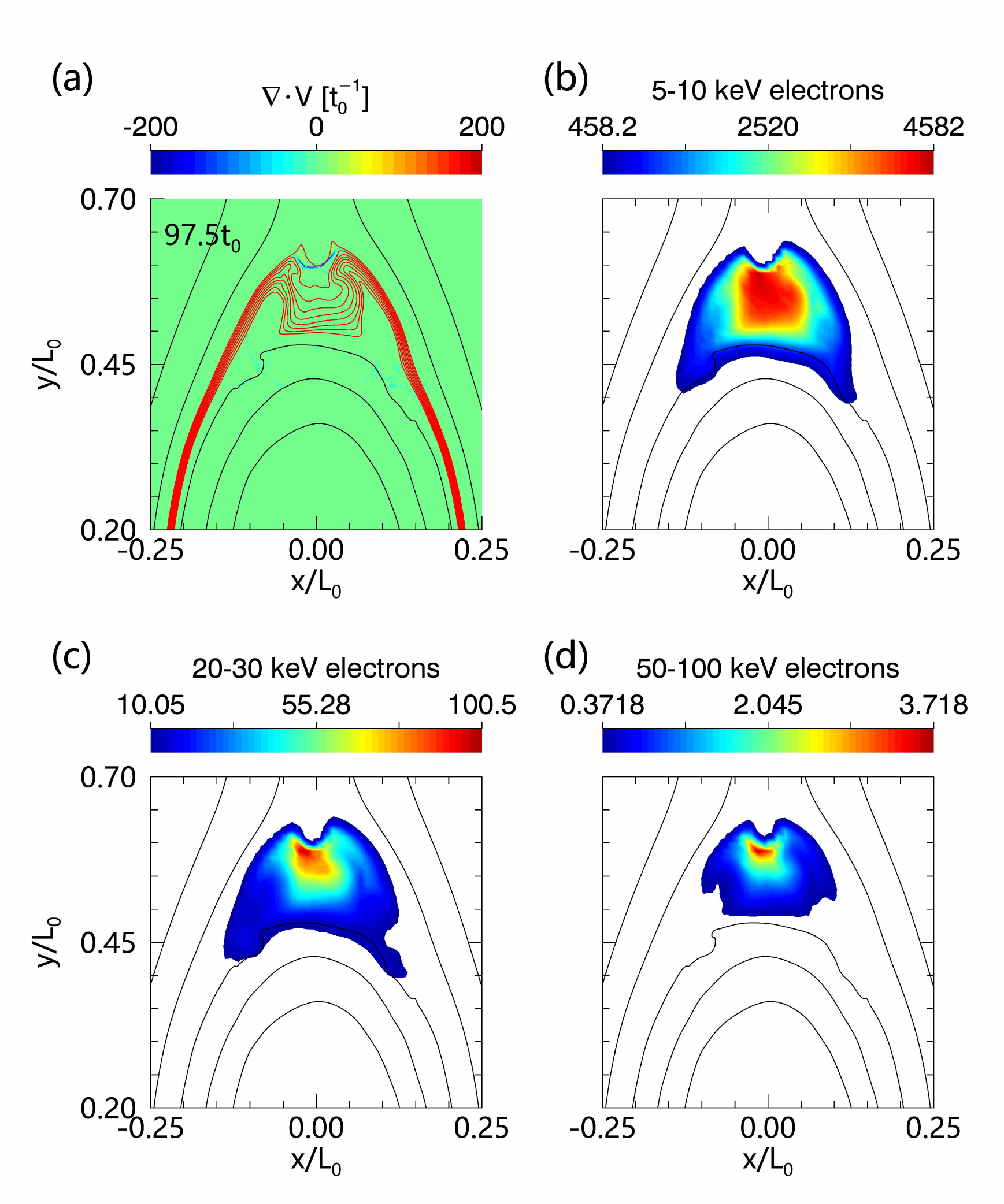}
\caption{
(a) Distributions of the divergence of plasma velocity $\nabla \cdot \textbf{V}$ at 97.5 $t_0$. The location of the termination shock is delineated by a narrow feature with negative $\nabla \cdot \textbf{V}$ values at $y \approx 0.65$. The red field lines illustrate the concave-downward magnetic trap structure at the loop top.
(b)--(d) Spatial distributions of accelerated electrons at 97.5 $t_0$ for three different energies, 5--10 keV, 20--30 keV, and 50--100 keV, respectively.
}
\label{fig:distribution}
\end{figure}

\begin{figure}
\centering
\includegraphics[width=0.9\linewidth]{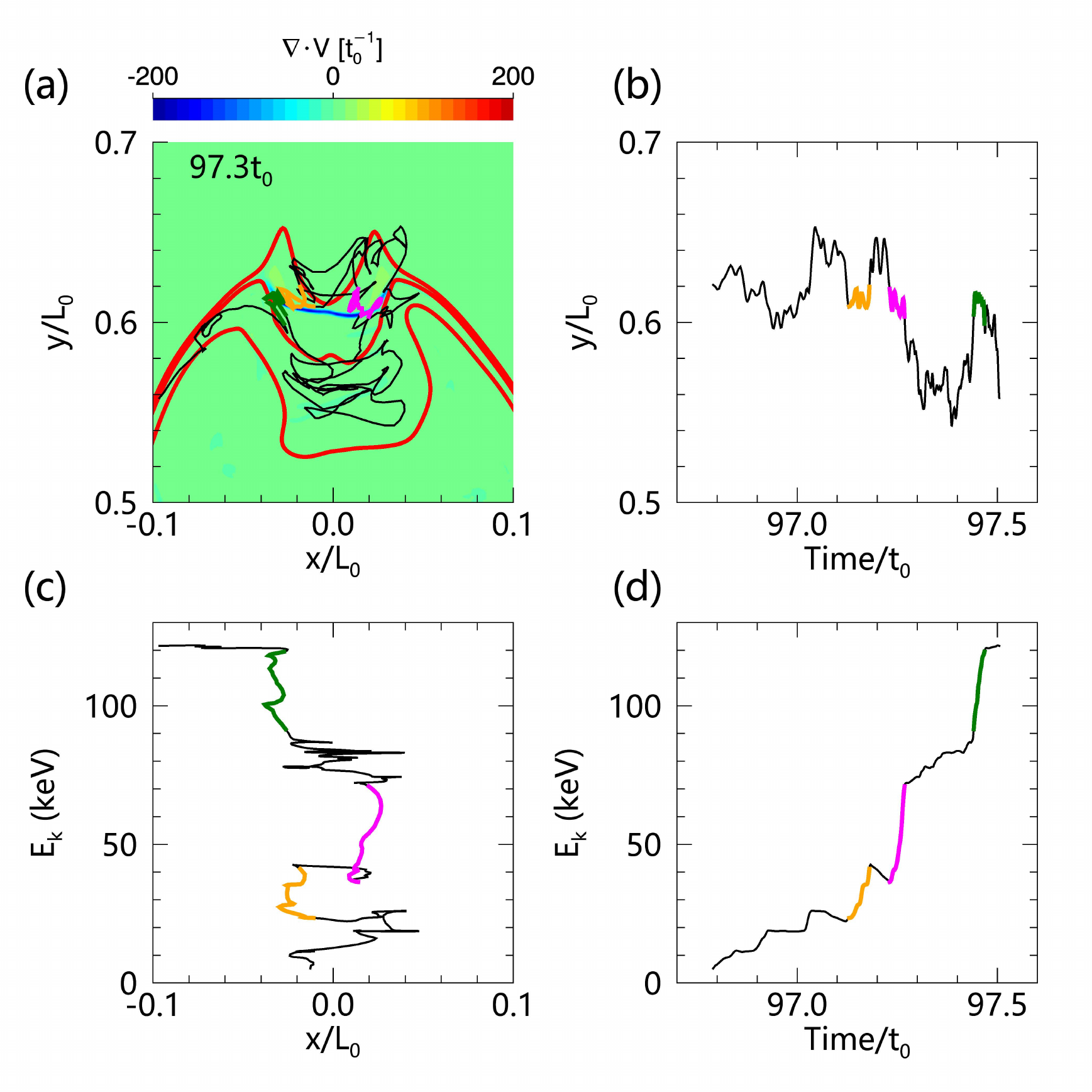}
\caption{
(a) Trajectory of a representative pseudo-particle overplotted on the $\nabla \cdot \textbf{V}$ map at 97.3 $t_0$. (b)--(d) The particle's $y$ position as a function of time, energy as a function of $x$ position and time.
Three rapid acceleration phases are highlighted in orange, magenta, and green, respectively.
}
\label{fig:trajectory}
\end{figure}

\begin{figure}
\centering
\includegraphics[width=0.9\linewidth]{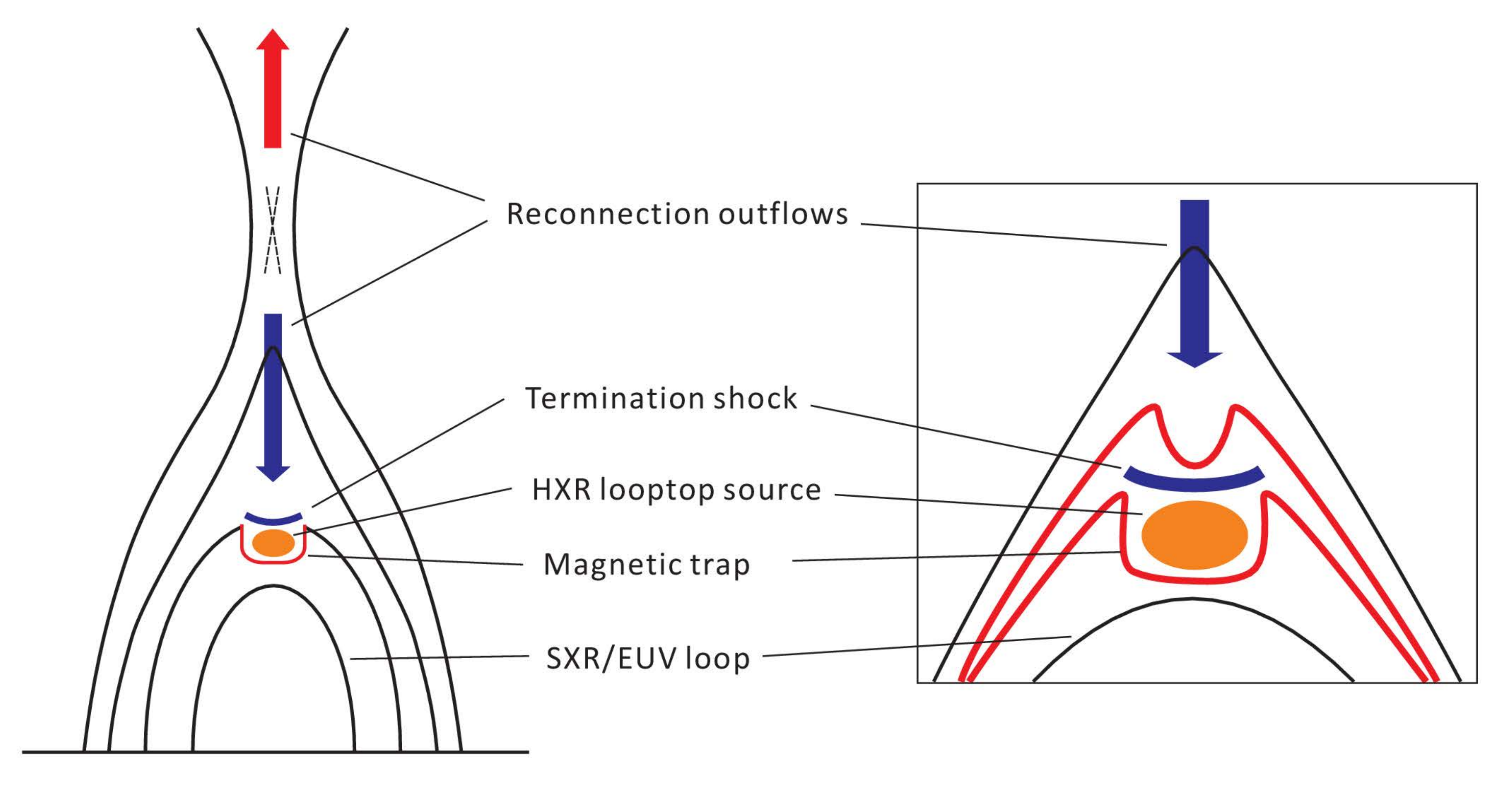}
\caption{
Schematic illustration of our model in the framework of the standard flare scenario for explaining the loop-top HXR sources by including a concave-downward magnetic trap structure at the loop-top. Energetic electrons are accelerated by the reconnection-driven termination shock and confined in the loop-top region by the magnetic trap structure where they produce the loop-top HXR emission.
}
\label{fig:cartoon}
\end{figure}


\end{document}